\documentclass[reprint,amsmath,amssymb,aps,prc,twocolumn,tightenlines,superscriptaddress]{revtex4-1}
\usepackage{graphicx}
\usepackage{units}
\usepackage{epsfig}
\usepackage{bm}
\usepackage{amsfonts}
\usepackage{amsmath}
\usepackage{amssymb}
\usepackage{makecell}
\usepackage[table]{xcolor}
\usepackage{gensymb}
\usepackage[utf8]{inputenc}

\newcommand{\pic}[2][1.0]{\includegraphics[width=#1\columnwidth]{#2}}

\newcommand{\nuc}[2]{\hbox{$^{#1}$#2}}

\begin{document}

\preprint{V. 0.5}

\title{Two-neutron knockout as a probe of the composition of states in \nuc{22}Mg, \nuc{23}Al, and \nuc{24}Si}


\author{B.~Longfellow}
\affiliation{National Superconducting Cyclotron Laboratory, Michigan State University, East Lansing, Michigan 48824, USA}
\affiliation{Department of Physics and Astronomy, Michigan State University, East Lansing, Michigan 48824, USA}
%
\author{A.~Gade}
\affiliation{National Superconducting Cyclotron Laboratory, Michigan State University, East Lansing, Michigan 48824, USA}
\affiliation{Department of Physics and Astronomy, Michigan State University, East Lansing, Michigan 48824, USA}
%
\author{J.~A.~Tostevin}
\affiliation{Department of Physics, Faculty of Engineering and Physical Sciences, University of Surrey, Guildford, Surrey GU2 7XH, UK}

\author{E.~C.~Simpson}
\affiliation{Department of Nuclear Physics, Research School of Physics, The Australian National University, Canberra ACT 2601, Australia}

\author{B.~A.~Brown}
\affiliation{National Superconducting Cyclotron Laboratory, Michigan State University, East Lansing, Michigan 48824, USA}
\affiliation{Department of Physics and Astronomy, Michigan State University, East Lansing, Michigan 48824, USA}

\author{A.~Magilligan}
\affiliation{National Superconducting Cyclotron Laboratory, Michigan State University, East Lansing, Michigan 48824, USA}
\affiliation{Department of Physics and Astronomy, Michigan State University, East Lansing, Michigan 48824, USA}

\author{D.~Bazin}
\affiliation{National Superconducting Cyclotron Laboratory, Michigan State University, East Lansing, Michigan 48824, USA}
\affiliation{Department of Physics and Astronomy, Michigan State University, East Lansing, Michigan 48824, USA}
\author{P.~C.~Bender}
\altaffiliation{Present address: Department of Physics, University of Massachusetts Lowell, Lowell, Massachusetts 01854, USA.}
\affiliation{National Superconducting Cyclotron Laboratory, Michigan State University, East Lansing, Michigan 48824, USA}
\author{M.~Bowry}
\altaffiliation{Present address: University of the West of Scotland, Paisley PA1 2BE, UK.}
\affiliation{National Superconducting Cyclotron Laboratory, Michigan State University, East Lansing, Michigan 48824, USA}

\author{B.~Elman}
\affiliation{National Superconducting Cyclotron Laboratory, Michigan State University, East Lansing, Michigan 48824, USA}
\affiliation{Department of Physics and Astronomy, Michigan State University, East Lansing, Michigan 48824, USA}

\author{E.~Lunderberg}
\affiliation{National Superconducting Cyclotron Laboratory, Michigan State University, East Lansing, Michigan 48824, USA}
\affiliation{Department of Physics and Astronomy, Michigan State University, East Lansing, Michigan 48824, USA}

\author{D.~Rhodes}
\affiliation{National Superconducting Cyclotron Laboratory, Michigan State University, East Lansing, Michigan 48824, USA}
\affiliation{Department of Physics and Astronomy, Michigan State University, East Lansing, Michigan 48824, USA}

\author{M.~Spieker}
\altaffiliation{Present address: Department of Physics, Florida State University, Tallahassee, Florida 32306, USA.}
\affiliation{National Superconducting Cyclotron Laboratory, Michigan State University, East Lansing, Michigan 48824, USA}

\author{D.~Weisshaar}
\affiliation{National Superconducting Cyclotron Laboratory, Michigan State University, East Lansing, Michigan 48824, USA}
\author{S.~J.~Williams}
\altaffiliation{Present address: Diamond Light Source, Harwell Science and Innovation Campus, Didcot, Oxfordshire, OX11 0DE, UK.}
\affiliation{National Superconducting Cyclotron Laboratory, Michigan State University, East Lansing, Michigan 48824, USA}
%

\date{\today}

\begin{abstract}

Simpson and Tostevin proposed that the width and shape of exclusive parallel
momentum distributions of the $A-2$ residue in direct two-nucleon knockout
reactions carry a measurable sensitivity to the nucleon
single-particle configurations and their couplings within the wave functions of
exotic nuclei. We report here on the first benchmarks and use of this new
spectroscopic tool. Exclusive parallel momentum distributions for states in the
neutron-deficient nuclei \nuc{22}Mg, \nuc{23}Al, and \nuc{24}Si populated in
such direct two-neutron removal reactions were extracted and compared to
predictions combining eikonal reaction theory and shell-model calculations. For
the well-known \nuc{22}{Mg} and \nuc{23}{Al} nuclei, measurements and calculations
were found to agree, supporting the dependence of the parallel momentum distribution
width on the angular momentum composition of the shell-model two-neutron amplitudes.
In \nuc{24}{Si}, a level at 3439(9)~keV, of relevance for the important \nuc{23}Al(p,$
\gamma$)\nuc{24}Si astrophysical reaction rate, was confirmed to be the $2_2^+$ state,
while the $4_1^+$ state, expected to be strongly populated in two-neutron knockout,
was not observed. This puzzle is resolved by theoretical considerations of the
Thomas-Ehrman shift, which also suggest that a previously reported 3471-keV state
in \nuc{24}Si is in fact the ($0_2^+$) level with one of the largest experimental
mirror-energy shifts ever observed.
\end{abstract}


\maketitle

One of the major endeavors in nuclear science is the exploration of the
evolution of nuclear structure far beyond the valley of $\beta$
stability. For years, direct one-nucleon knockout reactions from
projectiles at intermediate energies have been key tools in successfully
tracking changes in single-particle energies and strengths toward the nucleon
drip lines \cite{HANSEN200353,GADE200860,OBERTELLI2016131}. More recently,
it has been shown that two-proton and two-neutron removal from neutron-rich
and neutron-deficient projectiles, respectively, also proceed as direct
reactions \cite{BAZIN200391,YONEDA200674}.

By combining an eikonal model of the reaction dynamics, that assumes a sudden,
single-step removal of two nucleons, and shell-model calculations of the
two-nucleon amplitudes (TNAs), the cross sections for two-nucleon knockout
from the parent-nucleus ground state to each of the final states in the daughter
nucleus can be calculated \cite{TOSTEVIN200674}. Previous work has shown that
the shape of the parallel momentum ($p_{\|}$) distribution of the two-nucleon
knockout residues depends strongly on the total angular momentum $I$ of the
two removed nucleons, allowing spin values to be assigned to populated final
states \cite{SIMPSON2009102,SIMPSON200979,SANTIAGO-GONZALEZ201183}. One step
further, it was proposed that, since the two-nucleon overlaps contain components
with different values of total orbital angular momentum $\vec{L} = \vec{l}_1 +
\vec{l}_2$, information beyond the $I$ value can be probed. This opens up the
possibility to uniquely explore this composition and couplings within the wave
functions of rare isotopes \cite{SIMPSON201082}.

In the present work, this configuration sensitivity of the two-neutron
knockout-residue $p_{\|}$ distributions is explored with three $sd$-shell
cases where the incoming projectiles each have 12 neutrons:
\nuc{9}Be(\nuc{24}Mg,\nuc{22}Mg+$\gamma$)$X$,
\nuc{9}Be(\nuc{25}Al,\nuc{23}Al+$\gamma$)$X$, and
\nuc{9}Be(\nuc{26}Si,\nuc{24}Si+$\gamma$)$X$. From analysis of the exclusive
$p_{\|}$ distributions in two-neutron knockout, $J^{\pi}$ values are assigned
and the dependence of the width on the $L$ composition of the shell-model TNAs
is explored, demonstrating the significant utility of this reaction as a
spectroscopic tool.

The low-lying level scheme of
\nuc{22}Mg is well known \cite{ENSDF}, allowing comparisons of the widths of $p_{\|}$
distributions for several states of the same spin. In \nuc{23}Al,
only one excited state decays by $\gamma$-ray emission, a
core-coupled $7/2^+$ state at 1616(8)~keV \cite{GADE2008666}, facilitating clean
extraction of the exclusive $p_{\|}$ distributions for the two bound states.

Excitation energies and $J^{\pi}$ values in \nuc{24}Si are critical
for the \nuc{23}Al(p,$\gamma$) rate, which has significant impact on energy generation in Type-I X-ray
bursts \cite{PARIKH2008178,CYBURT2016830}. Energy values differing by several 10~keV were reported
originally~\cite{SCHATZ199779,YONEDA200674} and only recently, a
$d$(\nuc{23}Al,\nuc{24}Si$+\gamma$)$n$ measurement resolved the discrepancy and, in
addition to
states at 1874(3) and 3449(5)~keV, suggested a new level at 3471(6)~keV to be either the $0_2^+$ state
with an extremely large Thomas-Ehrman shift \cite{EHRMAN195181,THOMAS195288} or the $4_1^+$ state
\cite{WOLF2019122}. To date, all $J^{\pi}$ assignments are reported as
tentative. In this work, two-neutron knockout is used to assign $J^{\pi}$ values in this key
nucleus for the first time.

The experiment was performed at the National
Superconducting Cyclotron Laboratory \cite{GADE2016053003}. A secondary
beam including \nuc{24}Mg (54.5\%), \nuc{25}Al (29.5\%), and \nuc{26}Si (13.5\%)
was produced by impinging the 150~MeV/u \nuc{36}Ar primary beam on a
550~mg/cm$^2$ \nuc{9}Be target at the midacceptance position of the A1900
fragment separator \cite{MORRISSEY200390}. A 250~mg/cm$^2$ achromatic Al wedge
was used for secondary beam purification.

Two-neutron knockout reactions were induced on a 287(3)~mg/cm$^2$ \nuc{9}Be target in front of the S800 spectrograph \cite{BAZIN2003629}. The mid-target
energies for \nuc{24}Mg, \nuc{25}Al, and \nuc{26}Si were 95~MeV/u, 102~MeV/u,
and 109~MeV/u, respectively. Event-by-event identification of the incoming
projectiles and outgoing reaction products was performed using plastic timing
scintillators and the S800 focal-plane detectors
\cite{YURKON1999291}. The particle identification plot for incoming \nuc{26}Si is shown in Fig.~1 of Refs.~\cite{LONGFELLOW201897,LONGFELLOW201999}. For each event, the $p_{\|}$ of the reaction residue at the target was determined using the magnetic
rigidity of the S800 spectrograph and the particle trajectory reconstructed from the position and angle measured in the S800 focal plane. To
optimize momentum resolution, the S800 analysis beamline was operated in
dispersion-matched mode. The target was surrounded by the high-efficiency, 192-element CAESium-iodide scintillator ARray (CAESAR) \cite{WEISSHAAR2010615} to tag populated excited states by their in-flight $\gamma$ decays.

\begin{figure}
        \begin{center}
            \pic{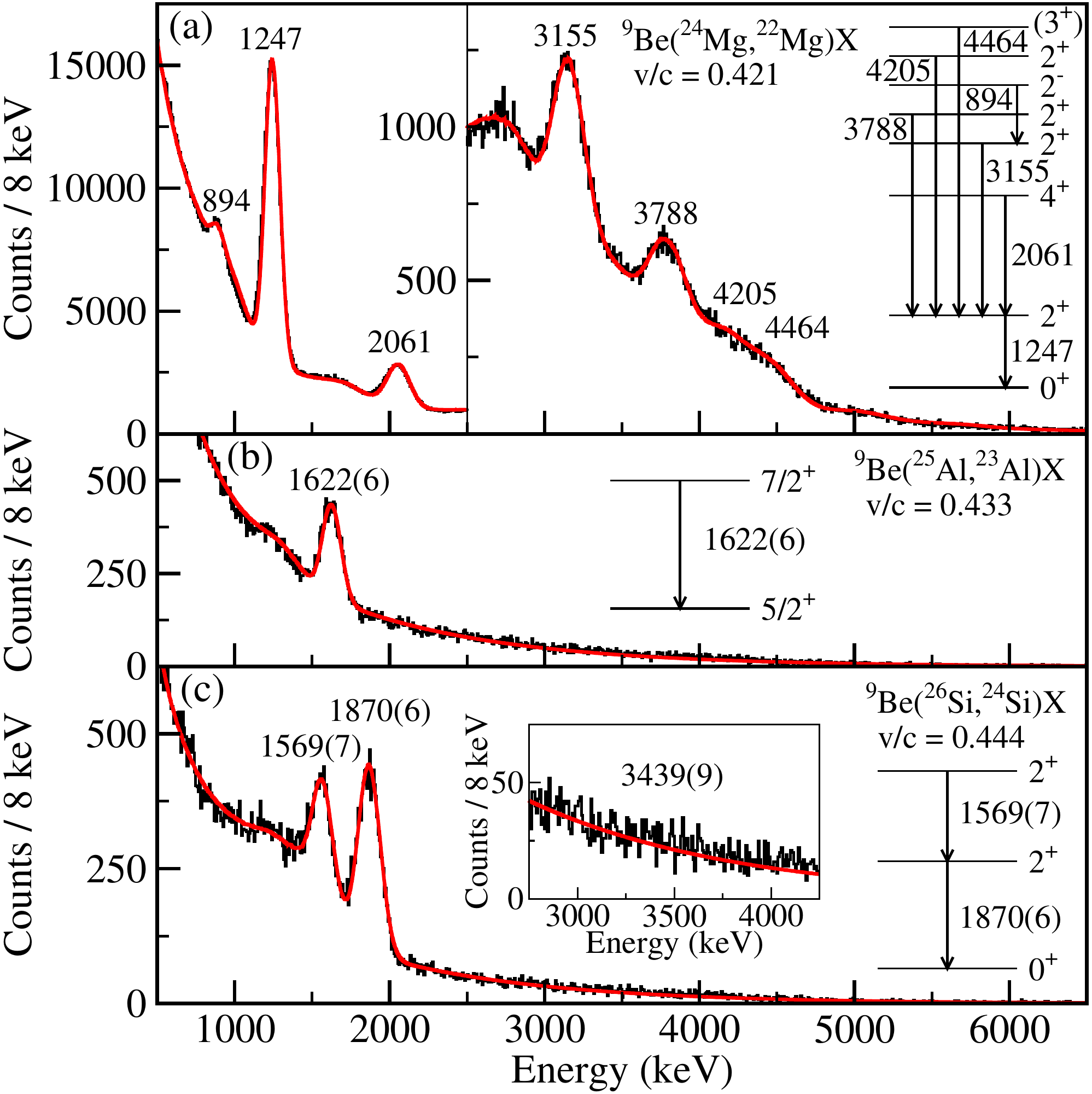}
            \caption{Doppler-corrected $\gamma$-ray spectra for the reactions
              \nuc{9}Be(\nuc{24}Mg,\nuc{22}Mg+$\gamma$)$X$
              (a), \nuc{9}Be(\nuc{25}Al,\nuc{23}Al+$\gamma$)$X$ (b), and
              \nuc{9}Be(\nuc{26}Si,\nuc{24}Si+$\gamma$)$X$ (c). The
              solid red curves are GEANT4 simulations of the
              observed transitions. Level schemes comprising the most
              intense transitions are displayed.
            }
            \label{fig:gam}
        \end{center}
\end{figure}

The Doppler-corrected $\gamma$-ray spectrum for \nuc{22}Mg produced from
two-neutron knockout is shown in Fig.~\ref{fig:gam}(a). The proton separation energy of \nuc{22}Mg is 5504.3(4)~keV \cite{ENSDF}. While peaks at 894, 1247, 2061, 3155, and 3788~keV are clearly visible, $\gamma$-ray transitions above 4~MeV are not resolved. To determine the
energies of possible transitions in this region, data from \nuc{24}Mg(p,t)\nuc{22}Mg \cite{CHAE200979,MATIC200980} and
\nuc{12}C(\nuc{12}C,2n+$\gamma$)\nuc{22}Mg \cite{SEWERYNIAK200594}, which also
result in the net loss of two neutrons, were utilized. In all cases, states
at 5452 and 5711~keV, which decay primarily by 4205 and 4464-keV $\gamma$ rays,
respectively, were populated. Consequently, transitions at these
energies were assumed in the fit. To determine exclusive cross
sections, the literature branching ratios of known weak decays from the states clearly observed
in this work were also included in the fit \cite{ENSDF}.

Fig.~\ref{fig:gam}(b) shows the Doppler-corrected $\gamma$ rays
detected in coincidence with \nuc{23}Al from two-neutron knockout. \nuc{23}Al has a low proton-decay threshold of 141.0(5)~keV \cite{ENSDF} and only one
$\gamma$-ray transition is visible at 1622(6)~keV, in agreement with Ref.~\cite{GADE2008666}.

Fig.~\ref{fig:gam}(c) displays the Doppler-corrected $\gamma$-ray
spectrum for \nuc{9}Be(\nuc{26}Si,\nuc{24}Si+$\gamma$)$X$. Clear peaks
at 1569(7) and 1870(6)~keV are visible. From $\gamma\gamma$ coincidences and intensities, the
1569-keV transition feeds the 1870-keV level. The energy for the first
excited state agrees within uncertainties with all previous
measurements \cite{SCHATZ199779,YONEDA200674,WOLF2019122}. The resulting energy of the 3439(9)-keV second-excited level, located just above the proton-emission threshold of
3293(20)~keV \cite{ENSDF} and of importance for the
\nuc{23}Al(p,$\gamma$)\nuc{24}Si rate, agrees with the
3441(10)-keV value from Ref.~\cite{SCHATZ199779}.

The $p_{\|}$ distributions for states in \nuc{22}Mg, \nuc{23}Al, and
\nuc{24}Si were obtained by gating on observed $\gamma$-ray transitions. The
distributions were background subtracted, with significant contributions from
Compton-scattered higher-energy transitions accounted for, and then corrected for
efficiency and feeding according to the level schemes in Fig.~\ref{fig:gam}. The ground-state $p_{\|}$ distributions were obtained by subtracting the distributions of direct feeders from the inclusive $p_{\|}$ distributions. The $p_{\|}$ distributions for the direct feeders in this subtraction were not feeding-corrected and therefore include contributions from higher-lying levels that do not $\gamma$ decay directly to the ground state.

The theoretical $p_{\|}$ distributions, calculated using eikonal reaction theory and
shell-model two-neutron amplitudes from the USD interaction
\cite{WILDENTHAL198411,BROWN198838}, were transformed to the laboratory frame, and
convoluted with a Heaviside function to account for reactions occurring at different
depths in the target.
To empirically model the low-momentum tails often observed in nucleon knockout, the
asymmetric $p_{\|}$ distributions of inelastically-scattered projectiles in coincidence
with $\gamma$ rays above 500~keV were folded with the calculated distributions following
the prescription of Ref.~\cite{STROBERG201490}. This approximates the kinematics of the
dissipative interactions with the target.

\begin{figure}
        \begin{center}
            \pic{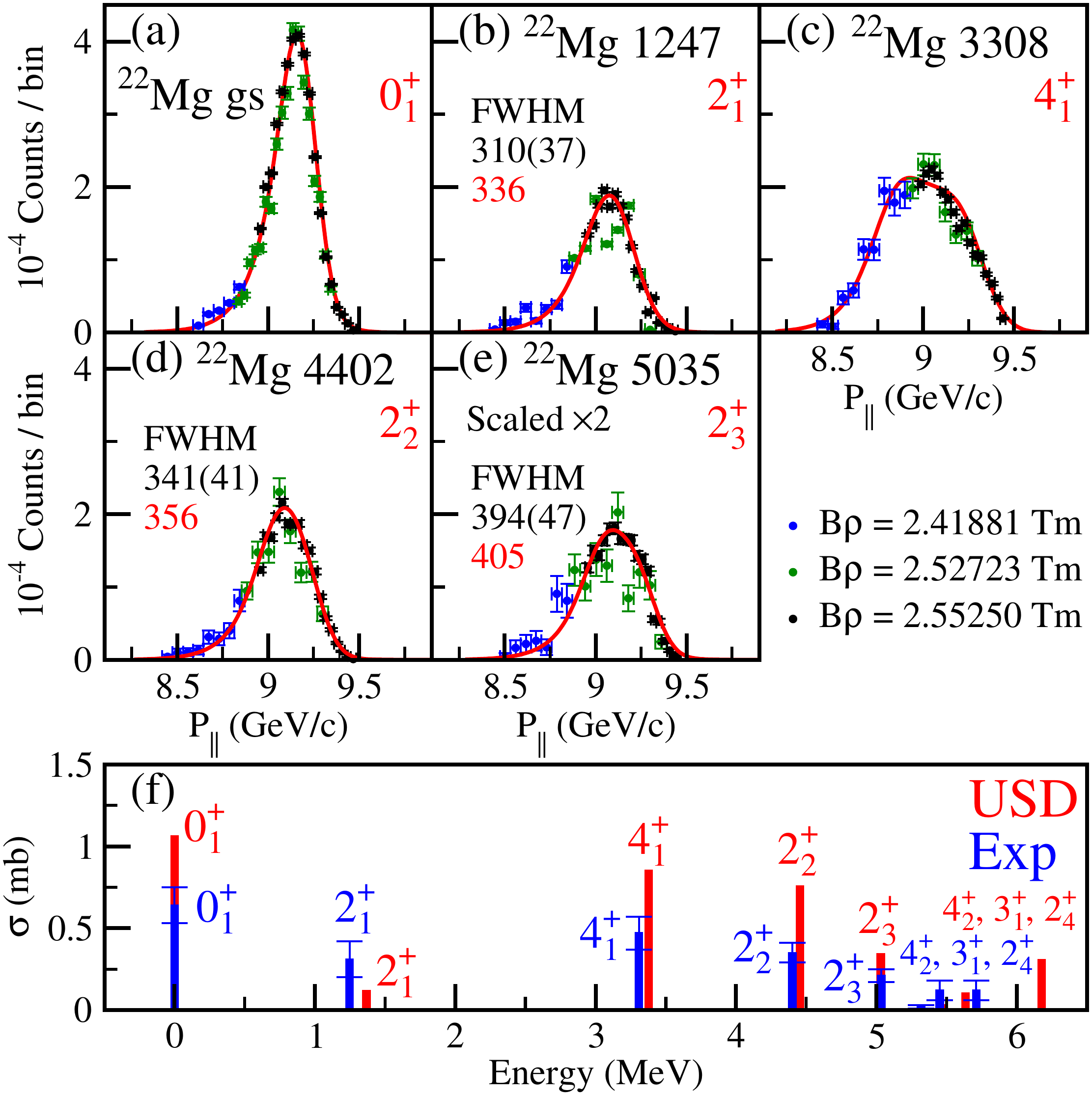}
            \caption{Parallel momentum ($p_{\|}$) distributions for states in
              \nuc{22}Mg populated in two-neutron knockout. The blue, green, and
              black points correspond to data taken at different magnetic
              rigidities (B$\rho$) of the S800 spectrograph. The vertical error bars are
              statistical. The solid red curves are the theoretical $p_{\|}$
              distributions scaled to best fit the
              data. The bottom panel compares the measured and
              calculated cross sections.
            }
            \label{fig:mg_p}
        \end{center}
\end{figure}

As seen in Fig.~\ref{fig:mg_p}, the shapes of the theoretical $p_{\|}$
distributions are in good agreement with the experimental results for the
previously-established $0_1^+$, $2_1^+$, $4_1^+$, $2_2^+$, and $2_3^+$ states in
\nuc{22}Mg. Experimental data were taken at several magnetic rigidity settings
of the S800 spectrograph to probe the full $p_{\|}$ distributions. Since two neutrons are knocked out from the $0^+$ ground state of \nuc{24}Mg, the total angular momentum $I$ of the removed neutrons is the spin $J$ of the populated state in \nuc{22}Mg. 
The spectroscopic power of two-neutron (2n) knockout is evident from the 
$2_1^+$, $2_2^+$, and $2_3^+$ state $p_{\|}$ distributions. Although of 
the same $J^{\pi}$, theory correctly predicts the observed variations in 
the widths of their $p_{\|}$ distributions, the result of different $L$ 
compositions of their TNAs \cite{SIMPSON201082}. That the three states 
are different is evident in Table~\ref{tab:tna}, where the largest $jj
$-coupled TNA for each final state involves different 2n configurations,
$[1d_{5/2},1d_{3/2}]$, $[1d_{5/2}]^2$, and $[1d_{5/2},2s_{1/2}]$. A full 
analysis of the $L$ makeup, from all TNAs (provided in the Supplemental Material \cite{supplement}), reveals a 
significant (narrower) $L=1$, $2_1^+$ component, that is smaller (relative 
to $L=2$) for $2_2^+$, and which is essentially zero for $2_3^+$, in line 
with the reported $p_{\|}$ distributions.



\begin{table}
\caption{TNAs calculated in the $jj$, np basis using the USD interaction for 
the first three $2^+$ states in \nuc{22}Mg populated in two-neutron knockout.}
\begin{ruledtabular}
\begin{tabular} {@{}lcccccc@{}}
J$^{\pi}$ & $[1d_{5/2}]^2$ & $[1d_{5/2},1d_{3/2}]$ & $[1d_{5/2},2s_{1/2}]$ & $[1d_{3/2}]^2$ &  $[1d_{3/2},2s_{1/2}]$  \\
\hline
\noalign{\vskip 1mm}
$2_1^+$	& -0.088 &  0.354 & -0.070 &  0.026 & -0.033 \\
$2_2^+$	& -0.756 &  0.222 & -0.219 & -0.044 &  0.221 \\
$2_3^+$	& -0.244 & -0.187 &	-0.377 & -0.169 &  0.117 \\
\end{tabular}
\end{ruledtabular}
\label{tab:tna}
\end{table}

Fig.~\ref{fig:mg_p}(f) shows the experimental partial cross
sections extracted for states in \nuc{22}Mg compared to the cross sections
calculated using the USD TNAs. Previous work on two-nucleon
knockout \cite{BAZIN200391,YONEDA200674,TOSTEVIN200674,WIMMER201285} has found a
ratio of approximately 0.5 between experimental and theoretical inclusive
cross sections in the $sd$-shell. For the $0_1^+$, $4_1^+$, $2_2^+$, and $2_3^+$
states, the partial cross sections are 0.64(11)~mb, 0.47(10)~mb, 0.35(6)~mb, and
0.21(4)~mb, giving ratios to theory of 0.60(10), 0.55(12), 0.46(8), and
0.62(11), respectively. Interestingly, the experimental cross section for the
$2_1^+$ state is 0.31(11)~mb while the theoretical prediction is 0.117~mb. This
is likely due to incomplete subtraction of feeding from several higher-lying
states, including $2^-$ and $3^-$ levels formed in the removal of one neutron
each from the $1d_{5/2}$ and $1p_{1/2}$ orbitals. Evidence
for their population here is, for example, the 894-keV $\gamma$ ray attributed
to the $2^- \rightarrow 2_2^+$ transition in \nuc{22}Mg
\cite{SEWERYNIAK200594}. From the mirror \nuc{22}Ne~\cite{ENSDF}, sizable
transitions to the
$2^+_1$ state, falling into the region of unresolved
transitions above 4~MeV, are expected from this $2^-$ state and from a $3^-$ state around
6~MeV. A partial cross section of 0.08(3)~mb to the $2^-$ was
inferred using only the 894-keV
transition but the total possible cross section to all $2^-$ states from
removal of one $1d_{5/2}$ and one $1p_{1/2}$ neutron is 1.687~mb. Only a small
fraction of this strength is needed to account for the suspected unobserved
feeding of the $2^+_1$ state. Unfortunately, the shapes of the calculated $2^-$
and $2_1^+$
$p_{\|}$ distributions are too similar to serve as a
discriminator.

The inclusive cross section for two-neutron knockout is
2.24(34)~mb, excluding the cross section to the $2^-$ state. The theoretical
inclusive cross section for $sd$-shell states up to the $2_3^+$ level is 3.572~mb,
giving a ratio for experiment to theory of 0.63(10). While this ratio is
slightly above the typical ratio of about 0.5, should cross section to negative-parity states have been misattributed to $sd$-shell
states, the ratio would decrease.

\begin{figure}
        \begin{center}
\pic{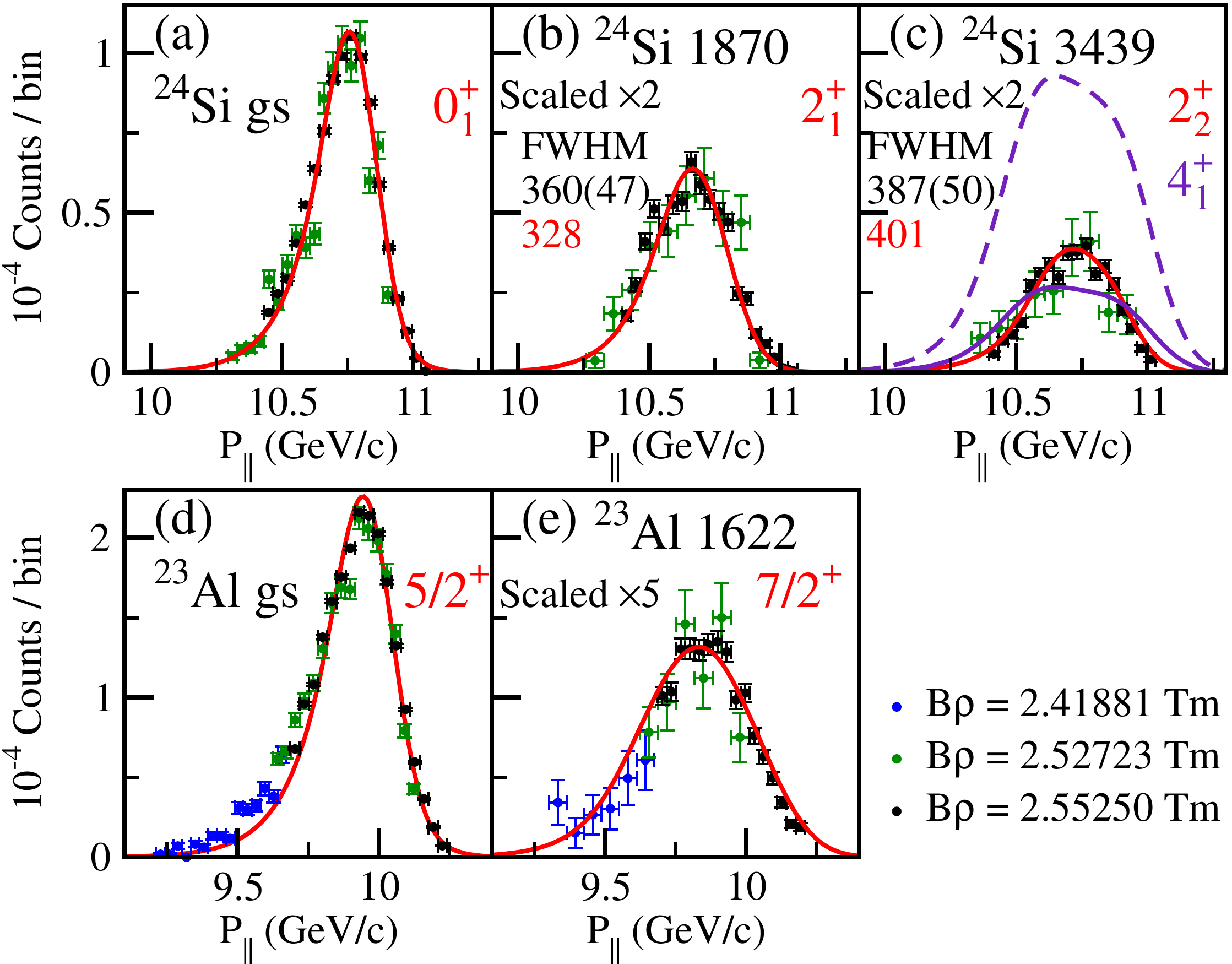}
            \caption{Parallel momentum ($p_{\|}$) distributions for states in \nuc{23}Al and \nuc{24}Si populated in two-neutron knockout. The blue, green, and black points correspond to data taken at different magnetic rigidities (B$\rho$) of the S800 spectrograph. The vertical error bars are statistical. The solid red and purple curves are the theoretical $p_{\|}$ distributions scaled to best fit the data. The dashed purple curve shows the distribution for the $4_1^+$ state in \nuc{24}Si assuming the theoretical cross section times 0.5.
            }
            \label{fig:al_si_p}
        \end{center}
\end{figure}

The measured and predicted $p_{\|}$ distributions for the $5/2^+$
ground  and $7/2^+$ excited state of \nuc{23}Al populated from the \nuc{25}Al($5/2^+)$ ground state are shown in
Fig.~\ref{fig:al_si_p}. Knockout to the $5/2^+$ level in \nuc{23}Al has
contributions from $sd$-shell neutrons coupled to $I$=0-4, with a predicted
dominance of the $I$=0 component. Knockout to the
$7/2^+$ state involves $I$=1-4 contributions, with $I$=4 larger than $I$=2 by about a
factor of two. For both, the odd $I$ TNAs are negligible. The experimental
$p_{\|}$ distributions
reflect this $I$ composition of the shell-model wave
function with a narrow $5/2^+$ and a broad $7/2^+$ $p_{\|}$ distribution.

The partial cross sections for the $5/2^+$ and
$7/2^+$ states are 0.60(8)~mb and 0.09(3)~mb, respectively, and the inclusive
cross section is 0.69(9)~mb.  The ratios to theory for the $5/2^+$, $7/2^+$, and
inclusive cross sections are 0.55(7), 0.54(18), and 0.55(7).
The centroids of the $p_{\|}$ distributions for the different final
states are slightly shifted with respect to each other, as reported in Ref.~\cite{SANTIAGO-GONZALEZ201183}.

Fig.~\ref{fig:al_si_p} shows the measured and calculated $p_{\|}$
distributions for levels in \nuc{24}Si populated in two-neutron knockout from
the \nuc{26}Si($0^+$) ground state. The shapes of the predicted distributions
for the ground and $2_1^+$ states agree well with the
data. Shell-model calculations and comparisons with the mirror
nucleus predict close-lying $2_2^+$ and $4_1^+$ levels in \nuc{24}Si. As seen in
Fig.~\ref{fig:al_si_p}(c), the data for the 3439(9)-keV level support a
$J^{\pi}$ assignment of $2_2^+$ rather than $4_1^+$. Since the experimental
$p_{\|}$ distribution is slightly narrower than the theoretical $2_2^+$ distribution, adding a $4_1^+$ component to the fit does not
improve the agreement. The measured widths of the $2_1^+$ and $2_2^+$ $p_{\|}$ distributions are consistent with the predicted dominance of $L$=1 and $L$=2, respectively, in the decomposition of the TNAs (see Tables III and IV and accompanying text of Ref.~\cite{SIMPSON201082}). In the mirror \nuc{24}Ne, the $2_2^+$ has relative
$\gamma$-decay intensities of 100.0(22) to the $2_1^+$
and 11.1(22) to the ground state. From our spectra (see Fig.~\ref{fig:gam}(c)), a ground-state branch of larger than 4\% can be excluded.

The partial cross sections for the $0_1^+$, $2_1^+$, and $2_2^+$ states in
\nuc{24}Si are 0.62(8)~mb, 0.17(3)~mb, and 0.13(3)~mb, giving ratios to theory
of 0.48(6), 0.53(9), and 0.40(9). The inclusive cross section is 0.92(10)~mb
giving a ratio to theory of 0.47(5). These results agree with the cross
sections reported in Ref.~\cite{YONEDA200674}.

In the recent $d$(\nuc{23}Al,\nuc{24}Si)$n$ work, $\gamma$-ray transitions at
1575(3)~keV from the ($2_2^+$) level at 3449(5)~keV
and 1597(5)~keV from the ($4_1^+$,$0_2^+$) level at 3471(6)~keV were proposed
\cite{WOLF2019122}. The results presented here for the
3439(9)-keV state confirm the $2_2^+$ assignment. If a transition at 1597~keV is
included in the fit of the $\gamma$-ray spectrum in Fig.~\ref{fig:gam}(c), its
intensity is at most 7\% of the 1570-keV transition, consistent with its
non-observation in Ref.~\cite{YONEDA200674}. The knockout calculation predicts a large
cross section of 0.935~mb to the
$4_1^+$ as compared to 0.329~mb to the $2_2^+$. The dashed purple curve in Fig.~\ref{fig:al_si_p}(c) shows the
expected $4^+_1$ $p_{\|}$ distribution assuming a cross
section of 0.5 times the prediction. If the 3471-keV level is the
$4_1^+$, then the 1597-keV transition should have been observed here. Conversely, the predicted cross section for the $0_2^+$ state in two-neutron knockout is only 0.005~mb, consistent with the non-observation of the 1597-keV transition. As noted
in Ref.~\cite{WOLF2019122}, if the 3471-keV level in
\nuc{24}Si is the $0_2^+$, then its energy is 1296~keV below the
$0_2^+$ state in \nuc{24}Ne.

To explore the expected Thomas-Ehrman (TE) shifts for states in \nuc{24}Si, proximity to the one-proton threshold of 3293(20)~keV \cite{ENSDF} was considered. For a state in \nuc{24}Si with excitation energy $E_x({\text{\nuc{24}Si}})$, the TE shift due to the one-proton separation energy relative to excited states in \nuc{23}Al below 4~MeV is:
\begin{equation*}
\begin{split}
    \textrm{TE}[E_x({\text{\nuc{24}Si}})] = \left( \frac{24}{23}\right)^2\sum_{ E_x({\text{\nuc{23}Al}})}^{4~\text{MeV}} C^2S(\text{\nuc{24}Si} \rightarrow \text{\nuc{23}Al}) \times \\
 \textrm{TE}_{WS}[S_p(\text{\nuc{24}Si}) + E_x({\text{\nuc{23}Al}})-E_x({\text{\nuc{24}Si}})].
\end{split}
\end{equation*}
Here, $\textrm{TE}_{WS}$ is the single-proton TE shift calculated from a
Woods-Saxon potential. The factor of $\left( \frac{24}{23}\right)^2$ is the
center-of-mass correction \cite{FRENCH61,DIEPERINK74}. The spectroscopic
factors, $C^2S$, are for one-proton 2$s_{1/2}$ overlaps as in
Ref.~\cite{LONGFELLOW201999}. The resulting relative TE shift for each level is
added to the measured energy of the \nuc{24}Ne mirror state. The results are
summarized in Table~\ref{tab:te}, together with the TE shift for the \nuc{23}Al $1/2^+$ state calculated using the
same method, in good agreement with experiment.

The TE shift for the $4^+$ state in \nuc{24}Si is minimal, predicting an energy of 4011~keV. If the $4^+$ level is around 4~MeV, then the one- and two-proton decays of the state would dominate, explaining the non-observation of its $\gamma$ decay in this work. The $0_2^+$ state in \nuc{24}Si is shifted down by 477 to 4290~keV. The $2s_{1/2}$ overlap that dominates the TE shift for the $0^+_2$ level is with the $1/2^+$
state in \nuc{23}Al, which itself has a large relative TE shift
of 426~keV.



The $0^+_2$ state has a large $[2s_{1/2}]^2$
two-proton overlap with the ground state of \nuc{22}Mg and the 3471(6)-keV level in \nuc{24}Si is only 37(20) keV above $S_{2p}=3434(19)$~keV \cite{ENSDF}.
Other examples of $0^+$ two-proton configurations
lying just above the two-proton
decay thresholds with large mirror-energy shifts
can be found in \nuc{18}Ne \cite{ENSDF} and \nuc{14}O \cite{CHARITY19}. Also, the unbound \nuc{26}O lies only 18~keV above the two-neutron separation
energy \cite{KONDO16}. If the 3471-keV level in \nuc{24}Si is our proposed $0^+_2$,
its large TE shift might be connected with its proximity
to the two-proton decay threshold.
Mirror symmetry is frequently evoked in nuclear astrophysics for
the identification and characterization of important levels for
capture-reaction networks, e.g. in the $rp$
process~\cite{RODRIGUEZ04,ROGERS11,LANGER14,MARGERIN15,SUZUKI17}. Isospin-symmetry breaking effects as large as the TE shift suggested here complicate such analyses significantly and must be considered.


\begin{table}
 \caption{Thomas-Ehrman (TE) shifts for states in \nuc{23}Al and \nuc{24}Si. The summed one-proton $\textrm{TE}$ contributions are added to the experimental energies of the mirror states in \nuc{23}Ne and \nuc{24}Ne \cite{ENSDF}. The $C^2S$ and $S_{p}'=S_p(\nuc{A}{Z})+E_x(\nuc{A-1}{Z-1})-E_x(\nuc{A}{Z})$ for the dominant term of the sum are shown. E$_{\text{mirr}}+\text{TE}$ energies are reported relative to the ground state and compared with the measured values of Ref.~\cite{ENSDF} for \nuc{23}Al and Ref.~\cite{WOLF2019122} for \nuc{24}Si. For the $4^+$ and $0^+_2$ states in \nuc{24}Si, E$_{\text{exp}}$ are both reported as 3471~keV for comparison.}
\begin{ruledtabular}
\begin{tabular} {@{}lccccccc@{}}
 & J$^{\pi}$ & $C^2S$ & $\textrm{S}_p'$ & $\textrm{TE}$ & E$_{\text{mirr}}$ & E$_{\text{mirr}}+\text{TE}$ & E$_{\text{exp}}$ \\
 & & & keV & keV & keV & keV & keV \\
\hline
\noalign{\vskip 1mm}
\nuc{23}Al	&	$5/2^+$	&	0.003	&	1484	&	-1	&	0	&	0	&	0	\\
	&	$1/2^+$	&	0.704	&	-409	&	-427	&	1017	&	590	&	550(20)	\\
\nuc{24}Si	&	$0^+$	&	0.226	&	3843	&	-57	&	0	&	0	&	0	\\
	&	$2^+$	&	0.191	&	1419	&	-84	&	1982	&	1955	&	1874(3)	\\
	&	$2^+$	&	0.500	&	-156	&	-253	&	3868	&	3672	&	3449(5)	\\
	&	$4^+$	&	0.052	&	3804	&	-17	&	3972	&	4011	&	3471(6)	\\
	&	$0^+$	&	1.083	&	-118	&	-534	&	4767	&	4290	&	3471(6)	\\

\end{tabular}
\end{ruledtabular}
\label{tab:te}
\end{table}


In summary, the reactions \nuc{9}Be(\nuc{24}Mg,\nuc{22}Mg+$\gamma$)$X$,
\nuc{9}Be(\nuc{25}Al,\nuc{23}Al+$\gamma$)$X$, and
\nuc{9}Be(\nuc{26}Si,\nuc{24}Si+$\gamma$)$X$ were used to benchmark the
sensitivity of theoretical parallel momentum distribution calculations to the
components in the shell-model two-neutron overlaps. In \nuc{22}Mg and
\nuc{23}Al, the shapes of the exclusive parallel momentum distributions were in
good agreement with theoretical predictions, realizing the high spectroscopic
potential of two-nucleon knockout. In \nuc{24}Si, the 3439-keV state, important
for the proton-capture reaction rate, was confirmed as the $2_2^+$
level. The predicted $4_1^+$ shell-model state in \nuc{24}Si,
expected to be strongly populated in two-neutron knockout, was not observed. By
considering Thomas-Ehrman shifts and proximity to the two-proton separation energy, we propose that the 3471-keV state reported in
Ref.~\cite{WOLF2019122} is the ($0_2^+$) rather than the $(4_1^+)$
state. Consequently, the experimental mirror-energy shift for the ($0_2^+$)
level in \nuc{24}Si is among the largest ever observed.

\begin{acknowledgments}
This work was supported by the National Science Foundation (NSF) under Grants No.\ PHY-1102511 and PHY-1565546, by the DOE National Nuclear Security Administration through the Nuclear Science and Security Consortium, under Award No.\ DE-NA0003180, and by the Department of Energy, Office of Nuclear Physics, under Grant No.\ DE-FG02-08ER41556 and DE-SC0020451. J.A.T acknowledges support from the Science and Technology Facilities Council (U.K.) Grant No.~ST/L005743/1. B.A.B. acknowledges support from NSF Grant No.\ PHY-1811855. The authors wish to thank Marek P{\l}oszajczak and Alexander Volya for helpful discussions.
\end{acknowledgments}

\end{document}